\documentclass[aps,prl,floatfix,twocolumn,bibnotes]{revtex4-1}
\usepackage{graphicx}
\usepackage{amsmath}
\usepackage{amssymb}
\usepackage{bm}
\usepackage{color}
\usepackage{dcolumn}
\usepackage{multirow}
\usepackage[caption=false]{subfig}
\usepackage{hyperref}

\begin{document}

\title{Prospects of a Pb$^{2+}$ ion clock}

\newcommand{\NIST}{
National Institute of Standards and Technology, Boulder, Colorado 80305, USA
}

\author{K. Beloy}
\email{kyle.beloy@nist.gov}
\affiliation{\NIST}

\date{\today}

\newcommand{\later}{\ensuremath{\spadesuit}}

\begin{abstract}
We propose a high-performance atomic clock based on the 1.81 PHz transition between the ground and first-excited state of doubly ionized lead. Utilizing an even isotope of lead, both clock states have $I=J=F=0$, where $I$, $J$, and $F$ are the conventional quantum numbers specifying nuclear, electronic, and total angular momentum, respectively. The clock states are nondegenerate and completely immune to nonscalar perturbations, including first order Zeeman and electric quadrupole shifts. Additionally, the proposed clock is relatively insusceptible to other frequency shifts (blackbody radiation, second order Zeeman, Doppler), accommodates ``magic'' rf trapping, and is robust against decoherence mechanisms that can otherwise limit clock stability. By driving the transition as a two-photon $E1$+$M1$ process, the accompanying probe Stark shift is appreciable yet manageable for practical Rabi frequencies.
\end{abstract}

\maketitle

Frequency standards referenced to transition frequencies in atomic systems (``atomic clocks'') are far and away the most accurate and precise metrological instruments ever developed~\cite{LudBoyYe15}. A recent measurement campaign between three distinct atomic clocks, for example, culminated in the ratios of the natural clock frequencies being determined at the part per quintillion ($10^{18}$) level, marking the lowest measurement uncertainty for any physical quantity to-date~\cite{BACON20}. The exquisite accuracy and precision afforded by atomic clocks can be leveraged to test physical theories, from foundational theories such as general relativity to more speculative theories attempting to explain dark matter~\cite{SafBudDeM18}. State-of-the-art optical atomic clocks interrogate either an ensemble of neutral atoms confined in an optical lattice~\cite{UshTakDas15,NemOhkTak16,McGZhaFas18,BotKedOel19,PizBreBar20} or a single ion confined in an rf Paul trap~\cite{MadDubZho12,HunSanLip16,HuaGuaBia17,BayGodJon18,BreCheHan19}. Regardless of the platform, realizing a high-performance atomic clock begins with a judicious choice of the atomic system and, more specifically, the atomic states to serve as the clock states.

Atomic states are specified, in part, by the conventional quantum numbers $I$, $J$, and $F$, which characterize the rotational symmetry (angular momentum) of the nuclear subsystem, the electronic subsystem, and the composite system, respectively. Meanwhile, interactions with external electromagnetic fields are conveniently decomposed into multipolar contributions, which have their own rotational symmetry. Namely, electric and magnetic $2^k$-pole (abbreviated $Ek$ and $Mk$, respectively) interactions are described by irreducible tensor operators of rank $k$. Moreover, higher-order perturbations involving these interactions can be cast in terms of irreducible tensor operators as well.
A familiar example is the second order $E1$ (Stark) shift, which can be partitioned into rank-0 (scalar), rank-1 (vector), and rank-2 tensor contributions. Even in absence of additional information about an atomic state, the rotational symmetries encapsulated by $I$, $J$, and $F$ provide valuable insight into the atom's susceptibility to external electromagnetic fields.

Optical ion clocks based on $^{40}$Ca$^+$~\cite{ChwBenKim09,MatHacLi12,ZhaHuaZha20}, $^{88}$Sr$^+$~\cite{MarBarHua04,MadDubZho12}, $^{138}$Ba$^+$~\cite{ArnKaeCha20}, and $^{172}$Yb$^+$~\cite{FurYehKal20} employ clock states with $I=0$ and $J\neq0$, with $F=J$ by consequence. $J\neq0$ implies that the electron cloud lacks spherical symmetry and, thus, is susceptible to nonscalar perturbations. In particular, these clocks must contend with considerable first order $M1$ (Zeeman) and $E2$ shifts. For the $^{88}$Sr$^+$ clock of Ref.~\cite{DubMadZho13}, for example, these shifts are of order $10^{-10}$ and $10^{-15}$, respectively (throughout, shifts and uncertainties given without units are understood to be in units of the respective clock frequency).

Optical ion clocks based on $^{27}$Al$^+$~\cite{BreCheHan19} and $^{115}$In$^+$~\cite{OhtLiNem20} employ clock states with $I\neq0$ and $J=0$, with $F=I$ by consequence. $J=0$ implies that the electron cloud possesses spherical symmetry and, thus, is insusceptible to nonscalar perturbations. Meanwhile, nonscalar nuclear perturbations are common-mode to both clock states, provided the nuclear substate is unchanged in the transition. However, this picture is incomplete. For $I\neq0$, the nucleus can be treated as a collection of multipole moments. Coupling of the electrons to the multipole fields of the nucleus (i.e., the hyperfine interaction) breaks the spherical symmetry of the electron cloud. This introduces susceptibility to nonscalar perturbations~\cite{LahMar75,ItaBerBru07,BelLeiIta17}. For the $^{27}$Al$^+$ clock of Ref.~\cite{BreCheHan19}, for example, the first order Zeeman shift is of order $10^{-12}$.

Optical ion clocks based on $^{199}$Hg$^+$~\cite{RosHumSch08}, $^{171}$Yb$^+$~\cite{GodNisJon14,HunSanLip16}, and $^{176}$Lu$^+$~\cite{ArnKaeRoy18} employ clock states with $I\neq0$ and $J\neq0$, with one or both clock states having $F\neq0$. As with the $J\neq0$ cases discussed above, these clocks must contend with considerable first order perturbations. Furthermore, each clock state is part of a hyperfine manifold. Higher-order perturbations, namely those involving even-parity interactions (e.g., $M1$, $E2$), are enhanced due to intramanifold couplings. For example, the $^{88}$Sr$^+$ clock of Ref.~\cite{DubMadZho13} and the $^{199}$Hg$^+$ clock of Ref.~\cite{RosHumSch08} both operate with a $\sim\!\mu$T bias magnetic field. The second order Zeeman shift is of order $10^{-20}$ for the former, whereas it is of order $10^{-15}$ for the latter.

The atomic states are $(2F+1)$-fold degenerate, corresponding to all possible values of the conventional quantum number $m_F$. In all cases above, at least one clock state has $F\neq0$ and is thus degenerate. In practice, a bias magnetic field is applied to lift the degeneracy and define the quantization axis. A sufficiently strong magnetic field must be applied to ensure the first order Zeeman effect dominates over other nonscalar perturbations, 
as well as to suppress line-pulling effects. Exploiting known dependencies on $m_F$ and external field parameters, several schemes have been devised in an effort to evade prominent nonscalar perturbations~\cite{Ita00,DubMadBer05,RooChwKim06,Bar15,ArnBar16,AhaSpeLer19,ShaAkeMan19,TanKaeArn19,
KaeTanArn20,LanHunSan20}. The standard approach involves interleaving distinct interrogation conditions (e.g., different $m_F$ substates or magnetic field directions~\cite{MadDubZho12,HunSanLip16,BreCheHan19}) and taking an appropriate average of the spectroscopic output. In any case, there is an operational burden associated with these schemes. Moreover, they are vulnerable to technical imperfections. In particular, uncontrolled variations in the applied magnetic field, the trapping potentials, or the stray-field environment can degrade either clock accuracy or stability.

In this Letter, we propose an ion clock employing $I=J=F=0$ clock states. In many regards, this is nature's ideal offering. The clock states are nondegenerate, completely immune to all nonscalar perturbations, and not part of a hyperfine manifold. This simplifies state preparation, eliminates the need to cancel nonscalar perturbations, and drastically relaxes the requirements for external field control. Note that these are highly appealing attributes, even before a system has been specified.

\begin{figure}[t]
\includegraphics[width=246pt]{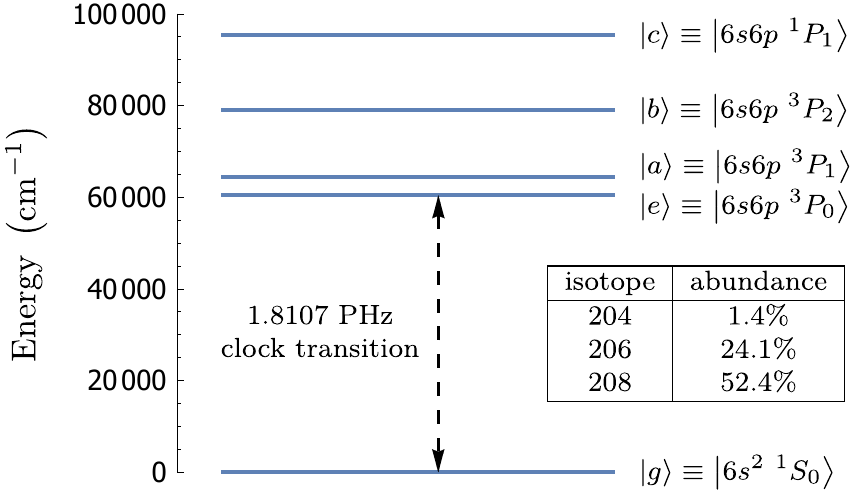}%
\caption{Lowest five states of the Pb$^{2+}$ energy spectrum. Conventional state labels are provided, as well as abbreviated labels used in this work. The clock transition is indicated. The $I=0$ isotopes (mass numbers) and their natural abundances are tabulated. All data is from Ref.~\cite{NISTASD}.}
\label{Fig:levelsplot}
\end{figure}

The criterion $I=0$ ($J=0$) implies an even number of nucleons (electrons). Only a handful of stable (or practically stable) odd-odd nuclides exist, none of which have $I=0$. Thus, satisfying both criteria requires a system with an even ionization degree. Doubly ionized group-14 ions have similar electronic structure to the singly ionized group-13 ions Al$^+$ and In$^+$, while offering $I=0$ isotopes (all the even isotopes). In this work, we focus on doubly ionized lead. The ionization energy of Pb$^+$ is 15.0 eV~\cite{NISTASD}, with laser ablation~\cite{AloCol07,CamSteChu09} or electron impact~\cite{DawParSmi00} being viable options for Pb$^{2+}$ production. Figure~\ref{Fig:levelsplot} presents the five lowest states of the Pb$^{2+}$ energy spectrum. For brevity, we label these states $\left|g\right\rangle$, $\left|e\right\rangle$, $\left|a\right\rangle$, $\left|b\right\rangle$, and $\left|c\right\rangle$. Conventional state labels are provided in Fig.~\ref{Fig:levelsplot} for reference. The proposed clock transition is between the ground state $\left|g\right\rangle\equiv\left|6s^2\,{^1S_0}\right\rangle$ and the first-excited state $\left|e\right\rangle\equiv\left|6s6p\,{^3P_0}\right\rangle$, with a transition frequency of 1.8107 PHz.

Two obstacles must be immediately addressed. The first is technical; the transition frequency lies outside the range readily covered by modern laser technology. The second is fundamental; angular considerations prohibit $F=0$ to $F=0$ transitions from proceeding via absorption or emission of a single photon. To simultaneously overcome both obstacles, we propose driving the transition as a two-photon process. From angular and parity considerations, the allowed channels are exclusively $Ek$+$Mk$ (i.e., the transition is mediated by one $Ek$ interaction and one $Mk$ interaction). All but the $E1$+$M1$ channel are prohibitively weak. Meanwhile, the $E1$+$M1$ channel cannot be driven with two photons from a single plane-wave probe field. This follows from the orthogonality of the electric and magnetic fields, together with the spherical symmetry of the clock states. In the following, we assume two probe fields, each with frequency approximately half the transition frequency. A small, well-defined frequency detuning between the probe fields avoids interference effects. In the interest of maximizing the Rabi frequency, we take the probe fields to be counterpropagating and orthogonally polarized~\cite{PbSM}. The probe frequency of 905 THz (wavelength of 331 nm) is within the means of modern laser technology.

The $E1$+$M1$ Rabi frequency is given by $\Omega=\left(8\pi/3\right)\left(\zeta/c\right)\sqrt{\mathcal{I}_1\mathcal{I}_2}\Lambda$, where $\zeta\equiv\left(2\varepsilon_0h c\right)^{-1}$, $\mathcal{I}_1$ and $\mathcal{I}_2$ are the probe intensities, and $\Lambda$ is an atomic factor~\cite{PbSM}. Here $\varepsilon_0$, $h$, and $c$ are the permittivity of free space, Planck's constant, and the speed of light, respectively. In terms of the dominant contributions,
\begin{gather}
\Lambda\approx
\left|\frac{
\left\langle e||\bm{\mu}||a\right\rangle
\left\langle a||\mathbf{D}||g\right\rangle
}{E_a-\frac{1}{2}\left(E_g+E_e\right)}
+
\frac{
\left\langle e||\bm{\mu}||c\right\rangle
\left\langle c||\mathbf{D}||g\right\rangle
}{E_c-\frac{1}{2}\left(E_g+E_e\right)}
\right|,
\label{Eq:Lambda}
\end{gather}
where $\left\langle i||\mathbf{D}||j\right\rangle$ and $\left\langle i||\bm{\mu}||j\right\rangle$ are reduced matrix elements of the conventional $E1$ and $M1$ operators, respectively, and $E_i$ are the atomic energies. In the nonrelativistic limit, $\left\langle a||\mathbf{D}||g\right\rangle$ and $\left\langle e||\bm{\mu}||c\right\rangle$ vanish due to spin selection rules. Thus, the $E1$+$M1$ channel opens up through relativistic effects. These effects enter in a largely correlated manner, such that the numerators in Eq.~(\ref{Eq:Lambda}) are approximately equal in magnitude but opposite in sign. This leads to a partial cancellation of the terms, the degree of which is largely determined by the relative size of the energy denominators (factor of $\approx\!2$). We use the CI+MBPT computational method~\cite{SavJoh02} to evaluate the matrix elements of Eq.~(\ref{Eq:Lambda}). The results are presented in Table~\ref{Tab:mels}. For the $E1$ matrix elements, our results are compared to the more complete CI+all-order results of Ref.~\cite{SafKozSaf12}. This provides a gauge for the accuracy of the intra- and intercombination matrix elements computed with the CI+MBPT method, which extends to the $M1$ matrix elements. This accuracy is more than sufficient for our purposes. Using the CI+MBPT matrix elements for consistency, together with experimental energies~\cite{NISTASD}, Eq.~(\ref{Eq:Lambda}) evaluates to $\Lambda\approx1.4$~a.u., where ``a.u.''\ stands for atomic units~\cite{aunote}. With this result, $\Omega$ can be evaluated for arbitrary probe intensities. Specific values are considered below.
The two-photon $E1$+$M1$ channel is also the dominant natural decay channel for the excited clock state. Using the CI+MBPT matrix elements, we calculate a natural lifetime of $\approx\!9\times10^{6}$ s~\cite{PbSM}.

\newcommand{\fnm}[1]{\footnotemark[#1]}
\begin{table}[tb]
\caption{Computed $E1$ and $M1$ matrix elements for Pb$^{2+}$ (a.u.). Present (CI+MBPT) and literature (CI+all-order) results are included.}
\label{Tab:mels}
\begin{ruledtabular}
\begin{tabular}{lcccc}
& $\left\langle g||\mathbf{D}||a\right\rangle$ 			
& $\left\langle g||\mathbf{D}||c\right\rangle$	
& $\left\langle e||\bm{\mu}||a\right\rangle$		
& $\left\langle e||\bm{\mu}||c\right\rangle$ \\
\hline\vspace{-3mm}\\
CI+MBPT								& 0.706			& 2.350 	& 0.674		& $-0.205$	\\
CI+all-order~\cite{SafKozSaf12}		& 0.644 		& 2.384
\end{tabular}
\end{ruledtabular}
\end{table}

Pb$^{2+}$ lacks an accessible $E1$ cycling transition. This motivates trapping a ``logic'' ion together with the Pb$^{2+}$ ``clock'' ion~\cite{SchRosLan05}. The logic ion enables sympathetic cooling of the clock ion, down to the ground state of the ions' coupled motion~\cite{CheBreCho17}. The logic ion further enables readout of the clock ion's internal state via quantum logic techniques. After detection, the Pb$^{2+}$ ion may be reset to $\left|g\right\rangle$ by illuminating it with 286 nm light, corresponding to the $M1$ transition $\left|e\right\rangle\rightarrow\left|c\right\rangle$. $\left|c\right\rangle$ decays to $\left|g\right\rangle$ with a lifetime of 0.3 ns~\cite{SafKozSaf12}. Alternatively, the detected state can simply be taken as the initial state for the subsequent interrogation. The alkali-like system Cd$^+$ is an intriguing logic ion candidate. In contrast to Pb$^{2+}$, the odd isotopes of Cd$^+$ are of interest (mass number 111 or 113, with $I=1/2$). The qubit states are furnished by the ground hyperfine doublet ($F=0,1$). Notably, Cd$^+$ has a charge-to-mass ratio similar to Pb$^{2+}$ (within 10\%), which promotes sympathetic cooling~\cite{KozSafCre18}. Meanwhile, two closed $E1$ cycling transitions are available (the alkali-like $D_1$ and $D_2$ transitions, with respective wavelengths of 227~nm and 215~nm). The viability of Cd$^+$ as a logic ion is supported by experimental work from different groups~\cite{DesHalLee04,MiaZhaSun15}.

We proceed to consider effects that may compromise accuracy or stability of the Pb$^{2+}$ clock. As noted above, Pb$^{2+}$ has similar electronic structure to Al$^+$. Currently, the Al$^+$ clock developed at NIST has the lowest systematic uncertainty of any atomic clock at $9.4\times10^{-19}$~\cite{BreCheHan19}. Moreover, the Al$^+$ clock employs quantum logic spectroscopy. For these reasons, the Al$^+$ clock is a particularly relevant case for comparison. In the following, we let $\nu$ represent the clock transition frequency and $\delta\nu$ (with appropriate subscript) a shift to this frequency.

The ubiquitous blackbody radiation (BBR) induces a Stark shift given by $\delta\nu_\mathrm{BBR}=-\zeta\,\mathcal{I}_\mathrm{BBR}\Delta\alpha$. Here $\mathcal{I}_\mathrm{BBR}\equiv4\sigma T^4$, where $\sigma$ is the Stefan-Boltzmann constant and $T$ is the BBR temperature. The atomic factor is $\Delta\alpha\equiv\alpha_e-\alpha_g$, where $\alpha_g$ and $\alpha_e$ are the static $E1$ polarizabilities of the ground and excited clock states, respectively. In the literature, we find the experimental value $\alpha_g=13.62(8)$~a.u.~\cite{HanKeeLun10} and the theoretical values $\alpha_g=13.3(3)$~a.u.\ and $\alpha_e=12.5(5)$~a.u.~\cite{SafKozSaf12}, where the numbers in parentheses specify standard uncertainty~\cite{uncnote}. Theoretical uncertainty in Ref.~\cite{SafKozSaf12} was estimated as (1 to 3)\% for $\alpha_g$ and (3 to 5)\% for $\alpha_e$; we have applied the median values of these respective estimates. Taking the weighted mean for $\alpha_g$ and the theoretical value for $\alpha_e$, we obtain $\Delta\alpha=-1.1(5)$~a.u., where a large relative uncertainty follows from a high degree of cancellation between the two polarizabilities. With this result, we find $\delta\nu_\mathrm{BBR}/\nu=5(2)\times10^{-18}$ at 300 K. This can be compared to Al$^+$, $\delta\nu_\mathrm{BBR}/\nu\approx-4\times10^{-18}$ at $300$~K~\cite{SafKozCla11,BreCheHan19}, where BBR insensitivity is generally regarded as one of the clock's greatest assets~\cite{RosItaSch06,LudBoyYe15}. Uncertainty in $\Delta\alpha$ can be reduced significantly. For Sr$^+$ and Ca$^+$ clocks, for instance, $\Delta\alpha$ has been evaluated to better than 0.2\%~\cite{DubMadTib14} and 0.03\%~\cite{HuaGuaZen19}, respectively. As this technique requires $\Delta\alpha<0$, it is applicable to Pb$^{2+}$ but not Al$^+$. Moreover, Ref.~\cite{DolBalNis15} argues that $T$ can be specified to within a fraction of a kelvin for room temperature ion clocks. We further find that dynamic~\cite{ItaLewWin82,PorDer06} and higher-multipolar~\cite{PorDer06} corrections enter at the $10^{-21}$ level at room temperature. We conclude that BBR shift uncertainty at the low-$10^{-20}$ level is feasible for a Pb$^{2+}$ clock operating at room temperature. Even lower uncertainty may be anticipated with cryogenic operation~\cite{RosHumSch08}.

We reiterate that the Pb$^{2+}$ clock has no first order Zeeman shift. The second order Zeeman shift is $\delta\nu_\mathrm{Zeeman}=-(1/2h)B^2\Delta\beta$, where $B$ is the root-mean-magnitude-squared value of the magnetic field. $B$ generally includes dc and ac contributions, with the latter being attributed to the rf trapping fields~\cite{BreCheBel19}. The atomic factor is $\Delta\beta\equiv\beta_e-\beta_g$, where $\beta_g$ and $\beta_e$ are the static $M1$ polarizabilities of the ground and excited clock states, respectively. $\Delta\beta$ is dominated by $\beta_e$, which is dominated by the coupling between $\left|e\right\rangle$ and $\left|a\right\rangle$. Specifically, $\Delta\beta\approx(2/3)\left|\left\langle e||\bm{\mu}||a\right\rangle\right|^2/\left(E_a-E_e\right)$, which evaluates to $\Delta\beta\approx17$~a.u. For a given $B$, $\delta\nu_\mathrm{Zeeman}/\nu$ is suppressed by more than a factor of 100 for Pb$^{2+}$ compared to Al$^+$. This suppression is due to the larger fine structure splitting ($66\times$) and the larger clock frequency ($1.6\times$), with the $M1$ matrix element being approximately equal for both systems. Furthermore, we reiterate that the Pb$^{2+}$ clock states are nondegenerate, eliminating the need for a bias magnetic field during clock spectroscopy. Supposing a finite value $B=2~\mu$T (consistent with dc and ac fields from the literature~\cite{DubMadZho13,BreCheBel19}), we find $\delta\nu_\mathrm{Zeeman}/\nu\approx-2.2\times10^{-21}$. Finally, we note that the logic ion can be exploited for magnetic field diagnostics~\cite{BreCheBel19}. For Cd$^+$, a $\sim\!\mu$T dc field induces a splitting of $\sim\!10^{-6}$ in the hyperfine clock frequency, which can be readily resolved~\cite{MiaZhaSun15}.

Ion motion introduces Doppler shifts. Despite strong ion confinement, ion clocks are susceptible to first order Doppler shifts. For instance, a first order Doppler shift of order $10^{-17}$ was observed in the NIST Al$^+$ clock, which was attributed to photoelectrons correlated with the clock cycle~\cite{BreCheHan19}. To mitigate this effect, the ion was alternately probed from opposite directions, with the spectroscopic signal being averaged. This approach assumes, e.g., invariance of the photoelectron emission dynamics over multiple interrogations. For the Pb$^{2+}$ clock, such averaging is unnecessary. By driving the two-photon transition with counterpropagating probe fields, the first order Doppler shift is coherently canceled each interrogation, irrespective of the ion motion. Note that a small frequency detuning between the probe fields (e.g., $\lesssim\!\text{MHz}$) is of negligible consequence.

Second order Doppler shifts must also be considered. Ion motion is conventionally partitioned into secular motion and micromotion~\cite{BerMilBer98}. Secular motion describes motion within a static ``pseudopotential.'' Micromotion further accounts for driven motion at the rf trapping frequency. 
In any case, the cumulative shift is $\delta\nu_\mathrm{Doppler}/\nu=-E_\mathrm{kinetic}/mc^2$, where $m$ is the mass of the clock ion and $E_\mathrm{kinetic}$ is its kinetic energy in the laboratory frame. The large mass of Pb$^{2+}$ provides a relative advantage. In particular, for similar trapping conditions (i.e., similar $E_\mathrm{kinetic}$), the shift is suppressed by a factor of $\approx\!8$ relative to Al$^+$.

The second order Doppler shift due to micromotion is the largest line item in the NIST Al$^+$ clock uncertainty budget~\cite{BreCheHan19}. While the mass of Pb$^{2+}$ offers a relative advantage, further advantage comes from the availability of a ``magic'' rf trap frequency $\nu_\mathrm{magic}$. When operating at $\nu_\mathrm{magic}$, the Doppler shift due to micromotion cancels with the correlated Stark shift~\cite{BerMilBer98,DubMadTib14,ArnKaeRoy18,HuaGuaZen19}. To lowest order, $\nu_\mathrm{magic}^2=-\left(h\nu/\Delta\alpha\right)(q/2\pi mc)^2$, where $q$ is the charge of the clock ion and $\Delta\alpha<0$ is a prerequisite. Using the result for $\Delta\alpha$ above, we conclude that a magic frequency exists in the range $80~\text{MHz}<\nu_\mathrm{magic}<250~\text{MHz}$ with $95\%$ confidence and with a most-probable value of $\nu_\mathrm{magic}\approx110~\text{MHz}$. We note that Pb$^{2+}$ strikes an excellent balance: $\left|\Delta\alpha\right|$ is small, providing suppressed Stark (including BBR) sensitivity, while $m$ is large, providing suppressed Doppler sensitivity and keeping $\nu_\mathrm{magic}$ within an experimentally convenient range.

The probe light induces a Stark shift according to $\delta\nu_\mathrm{probe}\approx-\zeta\left(\mathcal{I}_1+\mathcal{I}_2\right)\Delta\alpha$. More precisely, $\Delta\alpha$ should be replaced with a dynamic counterpart evaluated at the probe frequency. Using theoretical data tabulated in Ref.~\cite{SafKozSaf12}, the dynamic correction is found to be smaller than the uncertainty in $\Delta\alpha$, and we are content neglecting it. To provide a point of reference, we suppose $\mathcal{I}_1=\mathcal{I}_2=25~\text{kW}/\text{cm}^2$, which yields $\Omega/2\pi\approx16$~Hz and $\delta\nu_\mathrm{probe}/\nu\approx1\times10^{-12}$. While this shift is appreciable, we stress that it is only present during the probe pulses. Variants of Ramsey spectroscopy, which incorporate free evolution between probe pulses, have been developed to mitigate effects of probe-related shifts~\cite{YudTaiOat10,SanHunLan18}. In particular, auto-balanced Ramsey spectroscopy was recently demonstrated with a Yb$^+$ clock~\cite{SanHunLan18}. In that work, $\Omega/2\pi\approx17$~Hz and $\delta\nu_\mathrm{probe}/\nu\approx1\times10^{-12}$, comparable to the numbers above. Interrogation times $<\!100$~ms were used. With similar interrogation conditions for Pb$^{2+}$, we may expect similar probe Stark shift uncertainty to be attainable ($<\!10^{-18}$~\cite{SanHunLan18}). Moreover, given the inherent robustness of Pb$^{2+}$ against decoherence (see below), interrogation times $\gg\!100$~ms are feasible. This allows the size and influence of probe Stark shifts to be substantially reduced via lower probe intensities and greater fractional time dedicated to free evolution~\cite{YudTaiOat10}.

State-of-the-art ion clocks have instabilities attributed primarily to quantum projection noise (QPN)~\cite{ItaBerBol93}. QPN can be combated by increasing the interrogation time of the ion. As local oscillator (LO) technology progresses~\cite{HafFalGre15,CooRosLei15,MatLegHaf17,NorCliMun18,
RobOelMil19,OlsFoxFor19,LiuJagYu20}, longer interrogation times and unprecedented levels of ion clock stability can be pursued. However, decoherence of the ion's internal state sets a practical limit on the interrogation time and results in diminishing returns in clock stability as the interrogation time approaches the ion coherence time~\cite{PeiSchTam05}. Spontaneous emission is a fundamental decoherence mechanism, though technical processes may dominate. This is elucidated in Ref.~\cite{CleKimCui20}, where correlation spectroscopy was used to reject common-mode LO noise in a pair of Al$^+$ clocks. Using interrogation times up to 8 s, a coherence time consistent with the 20.6~s~\cite{RosSchHum07} natural lifetime of the excited clock state was observed. This required careful monitoring and rejection of magnetic field noise. Even still, it was necessary to invoke a combination of $m_F$ substates that suppresses first order Zeeman sensitivity by a factor of 15 relative to the stretched substates normally used in clock operation~\cite{BreCheHan19}. For Pb$^{2+}$, as previously noted, $\left|e\right\rangle$ decays on a timescale of a few months rather than tens of seconds. In principle, quenching mechanisms can shorten the lifetime. For room temperature operation, the BBR quenching rate is an order of magnitude lower than the spontaneous emission rate. For $\sim\!\mu$T dc magnetic fields, the Zeeman quenching rate is a few orders of magnitude smaller still. Moreover, as our analysis above attests, Pb$^{2+}$ is highly insensitive to external field fluctuations. In particular, we reiterate that the clock states are completely immune to nonscalar perturbations, including first order Zeeman and $E2$ shifts. Thus, we anticipate a Pb$^{2+}$ clock to be exceptionally robust against decoherence mechanisms that could otherwise restrict interrogation times and limit clock stability. The relatively high clock frequency provides additional leverage; compared to Al$^+$, Pb$^{2+}$ accommodates lower QPN with interrogation times half as long~\cite{ItaBerBol93}. Finally, we note that lead has three stable and two long-lived (half-life $>\!20$~y) $I=0$ isotopes. Using correlation spectroscopy, isotope shifts may be precisely measured using interrogation times far exceeding available LO coherence times~\cite{CleKimCui20}.

In conclusion, we propose an ion clock based on the 1.8107 PHz transition between the ground and first-excited state in an even isotope of Pb$^{2+}$. The intrinsic properties of this ion make it a promising candidate to realize unprecedented levels of ion clock accuracy and stability. Though the probe Stark shift may be appreciable, it can be managed using existing techniques~\cite{SanHunLan18}. Other doubly ionized group-14 ions may also be good candidates. Here we focused on Pb$^{2+}$, as second order Doppler and Zeeman shifts, the $E1$+$M1$ transition strength, and sensitivity to new physics~\cite{SafBudDeM18} all scale favorably with atomic number/weight. Finally, the attributes highlighted here in the context of a single-ion clock may lend themselves to broader applications, including multi-ion clock operation~\cite{ArnHajPae17,KelBurKal19,ShaAkeMan19,TanKaeArn19,KaeTanArn20} or ion-based quantum computing~\cite{BruChiMcC19}.

\begin{acknowledgments}
The author thanks S.~M.~Brewer, D.~R.~Leibrandt, and D.~B.~Hume for fruitful discussions and critical feedback. This work was supported by the National Institute of Standards and Technology/Physical Measurement Laboratory, an agency of the U.S. government, and is not subject to U.S. copyright.
\end{acknowledgments}


\begin{thebibliography}{77}
\bibitem [{aun()}]{aunote}%
  \BibitemOpen
  \href@noop {} {}\bibinfo {note} {We employ SI electromagnetic expressions.
  This is worth clarifying, as it affects the interpretation of ``a.u.''\ for
  the $M1$ matrix elements and quantities involving the $M1$ matrix
  elements.}\BibitemShut {Stop}%
\end{thebibliography}

%

\end{document}